\documentclass[11pt]{article}
\usepackage{natbib}
\usepackage{amssymb}
\newtheorem{defn}{Definition} 
\newtheorem{lem}{Lemma} 
\newtheorem{thm}{Theorem} 
\begin{document}

\title{Estimation of linear autoregressive models with Markov-switching,
the E.M. algorithm revisited. }

\author{Joseph Rynkiewicz\footnote{SAMOS/MATISSE, University of ParisI, 90 rue de Tolbiac, Paris, France, rynkiewi@univ-paris1.fr}}

\maketitle
\begin{abstract}
This work concerns estimation of linear autoregressive models with
Markov-switching using expectation maximisation (E.M.) algorithm.
Our method generalise the method introduced by Elliot for general
hidden Markov models and avoid to use backward recursion.
\end{abstract}

\section{Introduction}

In the present paper we consider an extension of basic (HMM). Let \( \left( X_{t},Y_{t}\right) _{t\in \mathbb Z} \) be the process such that 

\begin{enumerate}
\item \( \left( X_{t}\right) _{t\in \mathbb Z} \) is a Markov chain in
a finite state space \( \mathbb E=\{e_{1},...,e_{N}\} \), which can
be identified without loss of generality with the simplex of \( \mathbb R^{N} \),
where \( e_{i} \) are unit vector in \( \mathbb R^{N} \), with unity
as the \( i \)th element and zeros elsewhere. 
\item Given \( \left( X_{t}\right) _{t\in \mathbb Z} \), the process \( \left( Y_{t}\right) _{t\in \mathbb Z} \)
is a sequence of linear autoregressive model in \( \mathbb R \) and
the distribution of \( Y_{n} \) depends only of \( X_{n} \) and
\( Y_{n-1},\cdots ,Y_{n-p} \).
\end{enumerate}
Hence, for a fixed \( t \) , the dynamic of the model is : 

\(
Y_{t+1}=F_{X_{t+1}}(Y_{t-p+1}^{t})+\sigma _{X_{t+1}}\varepsilon _{t+1}\)
 with \( F_{X_{t+1}}\in \{F_{e_{1}},...,F_{e_{N}}\} \) linear functions,
\( \sigma _{X_{t+1}}\in \{\sigma _{e_{1}},...,\sigma _{e_{N}}\} \)
strictly positive numbers and \( \left( \varepsilon _{t}\right) _{t\in \mathbb N^{*}} \)
a i.i.d sequence of Gaussian random variable \( {\mathcal N}\left( 0,1\right)  \). 

\begin{defn}
Write \( {\mathcal F}_{t}=\sigma \left\{ X_{0},\cdots ,X_{t}\right\}  \),
for the \( \sigma  \)-field generated by \( X_{0},\cdots ,X_{t} \),
\( {\mathcal Y}_{t}=\sigma \left\{ Y_{0},\cdots ,Y_{t}\right\}  \), for
the \( \sigma  \)-field generated by \( Y_{0},\cdots ,Y_{t} \) and
\\
\( {\mathcal G}_{t}=\sigma \left\{ \left( X_{0},Y_{0}\right) ,\cdots ,\left( X_{t},Y_{t}\right) \right\}  \),
for the \( \sigma  \)-field generated by \( X_{0},\cdots ,X_{t} \)
and \( Y_{0},\cdots ,Y_{t} \).
\end{defn}
The Markov property implies here that \(
P\left( X_{t+1}=e_{i}\left| {\mathcal F}_{t}\right. \right) =P\left( X_{t+1}=e_{i}\left| X_{t}\right. \right) .\)
Write \(
a_{ij}=P\left( X_{t+1}=e_{i}\left| X_{t}=e_{j}\right. \right) \textrm{ and }A=\left( a_{ij}\right) \in \mathbb R^{N\times N}\)
 and define : \(
V_{t+1}:=X_{t+1}-E\left[ X_{t+1}\left| {\mathcal F}_{t}\right. \right] =X_{t+1}-AX_{t}.\)
 With the previous notations, we obtain the general equation of the
model, for \( t\in \mathbb N \)~:\begin{equation}
\label{modele_hybride}
\left\{ \begin{array}{c}
X_{t+1}=AX_{t}+V_{t+1}\\
Y_{t+1}=F_{X_{t+1}}(Y_{t-p+1}^{t})+\sigma _{X_{t+1}}\varepsilon _{t+1}
\end{array}\right. 
\end{equation}

The parameters of the model are the transition probabilities of the
matrix A, the coefficients of the linear functions \( F_{e_{i}} \)
and the variances \( \sigma _{e_{i}} \). A successfull method
for estimating such model is to compute the maximum likelihood estimator\footnote{%
This likelihood is computed conditionally to the first {}``p'' observations. 
} with the E.M. algorithm introduced by Demster , Lair and Rubin (1977).
Generally, this algorithm demands the calculus of the conditional
expectation of the hidden states knowing the observations (the E.-step),
this can be done with the Baum and Welch forward-backward algorithm (see Baum et al. (1970)).
The derivation of the M-step of the E.M. algorithm is then immediate since we can compute the optimal parameters of the regression functions thanks weighted linear regression. 

However we show here that we can also embed these two steps in only one. 
Namely we can compute, for each step of the E.M. algorithm, directly
the optimal coefficients of the regression functions as the variances
and the transition matrix thanks a generalisation of the method introduced
by \\Elliott (1994).

\section{Change of measure}

The fundamental technique employed throughout this paper is the discrete
time change of measure. Write \( \sigma  \) the vector \( (\sigma _{e_{1}},...,\sigma _{e_{N}}) \),
\( \phi (.) \) for the density of \( {\mathcal{N}}(0,1) \) and \( \left\langle .,.\right\rangle  \)
the inner product in \( \mathbb R^{N} \). 

We wish to introduce a new probability measure \( \bar{P} \), using
a density \( \Lambda  \), so that \( \frac{d\bar{P}}{dP}=\Lambda  \)
and under \( \bar{P} \) the random variables \( y_{t} \) are \( {\mathcal N}\left( 0,1\right)  \)
i.i.d. random variables. 

Define \[
\lambda _{l}=\frac{\left< \sigma ,X_{l-1}\right> \phi (y_{l})}{\phi (\varepsilon _{l})},  l\in \mathbb {N}^{*}, \mbox{ with }\Lambda _{0}=1 \mbox{ and }\Lambda _{t}=\prod _{l=1}^{t}\lambda _{l}\]

and construct a new probability measure \( \bar{P} \) by setting the restriction of the Radon-Nikodym derivative to \( {\mathcal{G}}_{t} \) equal
to \( \Lambda _{t} \). Then the following lemma is a straightforward adaptation
of lemma 4.1 of Elliot (1994) (see annexe).

\begin{lem}
Under \( \bar{P} \) the \( Y_{t} \) are \( {\mathcal N}\left( 0,1\right)  \)
i.i.d. random variables.
\end{lem}

Conversely, suppose we start with a probability measure \( \bar{P} \)
such that under \( \bar{P} \) 

\begin{enumerate}
\item \( \left( X_{t}\right) _{t\in \mathbb N} \) is a Markov chain with
transition matrix \( A \).
\item \( \left( Y_{t}\right) _{t\in \mathbb N} \) is a sequence of \( {\mathcal N}\left( 0,1\right)  \)
i.i.d. random variable.
\end{enumerate}
We construct a new probability measure \( P \) such that under \( P \) we have \\
\(
Y_{t+1}=F_{X_{t}}\left( Y_{t-p}^{t}\right) +\sigma _{X_{t}}\varepsilon _{t+1}.\)
 To construct \( P \) from \( \bar{P} \), we introduce\\ \( \bar{\lambda }_{l}:=\left( \lambda _{l}\right) ^{-1} \)
and \( \bar{\Lambda }_{t}:=\left( \Lambda _{t}\right) ^{-1} \) and
we define \( P \) by putting \( \left( \frac{dP}{d\bar{P}}\right) \left| _{{\mathcal G}_{t}}\right. =\bar{\Lambda }_{t} \), 

\begin{defn}
let \((H_{t}),\ t \in \mathbb{N}\) be a sequence adapted to \(({\mathcal{G}}_{t})\), We shall write~:
\[
\gamma_{t}(H_{t})=\bar {E}\left[\bar {\Lambda}_{t} H_{t}\left|{\mathcal{Y}}_{t}\right.\right] \mbox{ and }
\Gamma ^{i}\left( Y_{t+1}\right) =\frac{\phi \left( \frac{Y_{t+1}-F_{X_{t}}(Y_{t-p+1}^{t})}{\left\langle \sigma ,e_{i}\right\rangle }\right) }{\left\langle \sigma ,e_{i}\right\rangle \phi \left( Y_{t+1}\right) }.
\] 

\end{defn}
The proof of the following theorem is a detailled adaption of the proof of theorem 5.3 of Elliott (1994) (see annexe).

\begin{thm}
\label{theoreme_principal_elliott}Suppose \( H_{t} \) is a scalar
\( {\mathcal G} \)-adapted process of the form : \( H_{0} \) is \( {\mathcal F_{0}} \)
measurable, \( H_{t+1}=H_{t}+\alpha _{t+1}+\left\langle \beta _{t+1},V_{t+1}\right\rangle +\delta _{t+1}f\left( Y_{t+1}\right)  \),
\( k\geq 0 \), where \( V_{t+1}=X_{t+1}-AX_{t} \), \( f \) is a
scalar valued function and \( \alpha  \), \( \beta  \), \( \delta  \)
are \( {\mathcal G} \) predictable process (\( \beta  \) will be \( N \)-dimensional
vector process). Then : \begin{equation}
\label{equation_th_princip}
\begin{array}{lcl}
\displaystyle \gamma _{t+1}\left( H_{t+1}X_{t+1}\right)  & := & \gamma _{t+1,t+1}\left( H_{t+1}\right) \\
 & = & \sum _{i=1}^{N}\left\{ \left\langle \gamma _{t}\left( H_{t}X_{t}\right) ,\Gamma ^{i}\left( y_{t+1}\right) \right\rangle a_{i}\right. \\
 &  & +\gamma _{t}\left( \alpha _{t+1}\left\langle X_{t},\Gamma ^{i}\left( y_{t+1}\right) \right\rangle \right) a_{i}\\
 &  & +\gamma _{t}\left( \delta _{t+1}\left\langle X_{t},\Gamma ^{i}\left( y_{t+1}\right) \right\rangle \right) f\left( y_{t+1}\right) a_{i}\\
 &  & +\left( diag\left( a_{i}\right) -a_{i}a_{i}^{T}\right) \gamma _{t}\left( \beta _{t+1}\left\langle X_{t},\Gamma ^{i}\left( y_{t+1}\right) \right\rangle \right) 
\end{array}
\end{equation}
where \( a_{i}:=Ae_{i} \), \( a_{i}^{T} \) is the transpose of \( a_{i} \)
and \( diag\left( a_{i}\right)  \) is the matrix with vector \( a_{i} \)
for diagonal and zeros elsewhere. 
\end{thm}
We will now consider special cases of processes H. In all cases, we
will calculate the quantity \( \gamma _{t,t}\left( H_{t}\right)  \)
and deduce \( \gamma _{t}\left( H_{t}\right)  \) by summing the components
of \( \gamma _{t,t}\left( H_{t}\right)  \). Then, we deduce from the conditional Bayes' theorem the conditional expectation of \(H_t\)~:
\\ \( \hat{H}_{t}:=E\left[ H_{t}\left| {\mathcal Y}_{t}\right. \right] =\frac{\gamma _{t}\left( H_{t}\right) }{\gamma _{t}\left( 1\right) } \).
\section{Application to the Expectation (E.-step) of the E.M. algorithm }
We will use the previous theorem in order to compute conditional quantities
needed by the E.M. algorithm.

Let
\(
\displaystyle {\mathcal{J}}_{t}^{rs}=\sum _{l=1}^{t}\left< X_{l-1},e_{r}\right> \left< X_{l},e_{s}\right> \)
be the number of jump from state \( e_{r} \) to state \( e_{s} \)
at time \( t \), we obtain :\begin{equation}
\label{e_step_transition}
\begin{array}{lcl}
\displaystyle \gamma _{t+1,t+1}\left( {\mathcal{J}}_{t+1}^{rs}\right) = & \sum _{i=1}^{N} & \left< \gamma _{t,t}\left( {\mathcal{J}}_{t}^{rs}\right) ,\Gamma ^{i}\left( Y_{t+1}\right) \right> a_{i}\\
 & + & \left< \gamma _{t}\left( X_{t}\right) ,\Gamma ^{r}(Y_{t+1})\right> a_{sr}e_{s}.\\

\end{array}
\end{equation}
 Write now \( {\mathcal{O}}^{r}_{t} =\sum _{n=1}^{t+1}\left< X_{n},e_{r}\right>\) for the number of times, up
to \( t \), that \( X \) occupies the state \( e_{r} \).  We obtain \begin{equation}
\label{e_step_occupation}
\begin{array}{lcl}
\displaystyle \gamma _{t+1,t+1}\left( {\mathcal{O}}_{t+1}^{r}\right) = & \sum _{i=1}^{N} & \left< \gamma _{t,t}\left( {\mathcal{O}}_{t}^{r}\right) ,\Gamma ^{i}\left( Y_{t+1}\right) \right> a_{i}\\
 & + & \left< \gamma _{t}\left( X_{t}\right) ,\Gamma ^{r}(Y_{t+1})\right> a_{r}.
\end{array}
\end{equation}

For the regression functions, the M-Step of the E.M. algorithm is achieved by finding the parameters
minimising the weighted sum of squares : \[
\sum _{t=1}^{n}\gamma _{i}\left( t\right) \left( y_{t}-\left( a_{0}^{i}+a_{1}y_{t-1}+\cdots +a_{p}y_{t-p}\right) ^{2}\right) \]
where \( \gamma _{i}\left( t\right)  \) is the conditional expectation
of the hidden \( e_{i} \) at time \( t \) knowing the observations
\( y_{-p+1},\cdots ,y_{n} \). 

Write \( \psi ^{T}(t)=(1,y_{t-1},...,y_{t-p}) \) and \( \theta _{i}=(a_{0}^{i},...,a_{p}^{i}) \),
suppose that the matrix \( \left[ \sum _{t=1}^{n}\gamma _{i}\left( t\right) \psi (t)\psi ^{T}(t)\right]  \)
is invertible. The estimator \( \hat{\theta }_{i}(n) \)~of \( \theta _{i} \)
is given by :\[
\hat{\theta }_{i}(n)=\left[ \sum _{t=1}^{n}\gamma _{i}\left( t\right) \psi (t)\psi ^{T}(t)\right] ^{-1}\sum _{t=1}^{n}\gamma _{i}\left( t\right) \psi (t)Y_{t}.\]

Hence, in order to compute \( \hat{\theta }_{i}(n) \), we need to
estimate the conditional expectation of the following processes : 

\begin{enumerate}
\item \[
\displaystyle {\mathcal{TA}}_{t+1}^{r}(j)=\sum _{l=1}^{t+1}\left< X_{l},e_{r}\right> Y_{l-j}Y_{l+1}\]
 for \( -1\leq j\leq p \) and \( 1\leq r\leq N \). 
\item \[
\displaystyle {\mathcal{TB}}_{t+1}^{r}(i,j)=\sum _{l=1}^{t+1}\left< X_{l},e_{r}\right> Y_{l-j}Y_{l-i}\]
 for \( 0\leq j,i\leq p \) and \( 1\leq r\leq N \).
\item \[
\displaystyle {\mathcal{TC}}_{t+1}^{r}=\sum _{l=1}^{t+1}\left< X_{l},e_{r}\right> Y_{l+1}.\]
 
\item \[
\displaystyle {\mathcal{TD}}_{t+1}^{r}(j)=\sum _{l=1}^{t+1}\left< X_{l},e_{r}\right> Y_{l-j}\]
 for \( 0\leq j\leq p \) and \( 1\leq r\leq N \).
\end{enumerate}
Applying theorem (\ref{theoreme_principal_elliott}) with \( H_{t+1}(j)={\mathcal{TA}}_{t+1}^{r}(j) \),
\( H_{0}=0 \), \( \alpha _{t+1}=0 \), \( \beta _{t+1}=0 \), \( \delta _{t+1}=\left< X_{t},e_{r}\right> Y_{t-j} \)
and \( f(Y_{t+1})=Y_{t+1} \), if \( j\ne -1 \) or \( \delta _{t+1}=\left< X_{t},e_{r}\right>  \)
and \( f(Y_{t+1})=Y_{t+1}^{2} \) if \( j=-1 \), gives us\begin{equation}
\label{e_step_covariance}
\begin{array}{lcl}
\displaystyle \gamma _{t+1,t+1}\left( {\mathcal{TA}}_{t+1}^{r}(j)\right) = & \sum _{i=1}^{N} & \left< \gamma _{t,t}\left( {\mathcal{TA}}_{t}^{r}(j)\right) ,\Gamma ^{i}(Y_{t+1})\right> a_{i}\\
 & + & \left< \gamma _{t}(X_{t}),\Gamma ^{r}(Y_{t+1})\right> Y_{t-j}Y_{t+1}a_{r},
\end{array}
\end{equation}
 where \( a_{r} \) is the \( r \)-th column of \( A \). 

Then, applying theorem (\ref{theoreme_principal_elliott}) with

\( H_{t+1}(j)={\mathcal{TB}}_{t+1}^{r}(i,j) \), \( H_{0}=0 \), \( \alpha _{t+1}=0 \),
\( \beta _{t+1}=0 \) , \( \delta _{t+1}=\left< X_{t},e_{r}\right> Y_{t-j}Y_{t-i} \)
and \( f(Y_{t+1})=1 \) gives :\begin{equation}
\label{e_step_cov_passe}
\begin{array}{lcl}
\displaystyle \gamma _{t+1,t+1}\left( {\mathcal{TB}}_{t+1}^{r}(i,j)\right) = & \sum _{i=1}^{N} & \left< \gamma _{t,t}\left( {\mathcal{TB}}_{t}^{r}(j)\right) ,\Gamma ^{i}(Y_{t+1})\right> a_{i}\\
 & + & \left< \gamma _{t}(X_{t}),\Gamma ^{r}(Y_{t+1})\right> Y_{t-j}Y_{t-i}a_{r}.
\end{array}
\end{equation}

Next, applying theorem (\ref{theoreme_principal_elliott}) with

\( H_{t+1}={\mathcal{TC}}_{t+1}^{r} \), \( H_{0}=0 \), \( \alpha _{t+1}=0 \),
\( \beta _{t+1}=0 \), \( \delta _{t+1}=\left< X_{t},e_{r}\right>  \)
and \( f(Y_{t+1})=Y_{t+1} \) gives : \begin{equation}
\label{e_step_moyenne}
\begin{array}{lcl}
\displaystyle \gamma _{t+1,t+1}\left( {\mathcal{TC}}_{t+1}^{r}\right) = & \sum _{i=1}^{N} & \left< \gamma _{t,t}\left( {\mathcal{TC}}_{t}^{r}(j)\right) ,\Gamma ^{i}(Y_{t+1})\right> a_{i}\\
 & + & \left< \gamma _{t}(X_{t}),\Gamma ^{r}(Y_{t+1})\right> Y_{t+1}a_{r}.
\end{array}
\end{equation}

Finally, applying theorem (\ref{theoreme_principal_elliott}) with

\( H_{t+1}(j)={\mathcal{TD}}_{t+1}^{r}(j) \), \( H_{0}=0 \), \( \alpha _{t+1}=0 \),
\( \beta _{t+1}=0 \) , \( \delta _{t+1}=\left< X_{t},e_{r}\right> Y_{t-j} \)
and \( f(Y_{t+1})=1 \) gives : \begin{equation}
\label{e_step_moyenne_passe}
\begin{array}{lcl}
\displaystyle \gamma _{t+1,t+1}\left( {\mathcal{TD}}_{t+1}^{r}(j)\right) = & \sum _{i=1}^{N} & \left< \gamma _{t,t}\left( {\mathcal{TD}}_{t}^{r}(j)\right) ,\Gamma ^{i}(Y_{t+1})\right> a_{i}\\
 & + & \left< \gamma _{t}(X_{t}),\Gamma ^{r}(Y_{t+1})\right> Y_{t-j}a_{r}.
\end{array}
\end{equation}

The {}``Maximisation'' pass of the E.M. algorithm is now achieved
by updating the parameters in the following way.
\paragraph{Parameters of the transition matrix}
The parameter of the transition matrix will be updates with the formula~:
\begin{equation}
\label{m_step_transition}
\hat{a}_{sr}=\frac{\gamma _{T}\left( {\mathcal{J}}_{T}^{sr}\right) }{\gamma _{T}\left( {\mathcal{O}}_{T}^{r}\right) }.
\end{equation}
\paragraph{Parameters of the regression functions}
For \( 1\leq r\leq N \), let \\ \( R^{r}:=\left( R_{ij}^{r}\right) _{1\leq i,j\leq p+1} \)
be the symmetric with \\ \(
R_{11}^{r}=1,R_{1j}^{r}=R_{j1}^{r}=\hat{\mathcal{T}}\mathcal{D}^{r}(j)\mbox {,\, }R_{ij}=\hat{\mathcal{T}}\mathcal{B}^{r}{\left( i-1,j-1\right) }\)
 and \\ \(
C^{r}=(\hat{\mathcal{T}}\mathcal{C}^{r},(\hat{\mathcal{T}}\mathcal{A}^{r}(i))_{0\leq i\leq p})\)
 we can then compute the updated parameter \( \hat{\theta }_{r} \)
of the regression function \( F_{e_{r}} \) with the formula : \begin{equation}
\label{m_step_regression}
\hat{\theta }_{r}={\left(R^{r}\right)}^{-1}C^{r}
\end{equation}
\paragraph{Parameters of the variances}
Finally, thanks the previous conditional expectations, we can directly calculate
the parameters \( \hat{\sigma }_{1},...,\hat{\sigma }_{N} \), since for \( 1\leq r\leq N \) the conditional expectation of the mean square error of the rth model is \begin{equation}
\label{m_step_variance}
\hat{\sigma }_{r}^{2}=\frac{1}{{\mathcal{O}}_{r}}\left( \hat{\mathcal{T}}\mathcal{A}^{r}(-1)+\hat{\theta }^{T}_{r}R^{r}\hat{\theta }_{r}-2\hat{\theta }^{T}_{r}C^{r}\right) .
\end{equation}
 This complete the M-step of the E.M. algorithm. 
\section{conclusion}
Using the discrete Girsanov measure transform, we propose an new way to apply the E.M. algorithm in the case of Markov-switching linear autoregressions. 

Note that, contrary to the Baum and Welch algorithm, we don't use backward recurrence, altought the cost of calculus slighty increase since the number of operations is multiplicated by \(\frac{N}{2}\), where \(N\) is the number of hidden state of the Markov chain.
\section*{References}
Baum, L.E., Petrie, T., Soules, G. and Weiss N. A maximisation technique occuring in the statistical estimation of probabilistic functions of Markov processes. \emph{Annals of Mathematical statistics}, 41:1:164-171, 1970\\
\\
Demster, A.P., Lair N.M. and Rubin, D.B. (1977) Maximum likelihood
from incomplete data via the E.M. algorithm. \emph{Journal of the
Royal statistical society of London}, Series B:39:1--38, 1966.\\
\\
Elliott,R.J. (1994) Exact Adaptative Filters for Markov chains observed in Gaussian Noise \emph{Automatica} 30:9:1399-1408, 1994.\\

\section*{Annexe}

\subsection*{Proof of lemma 1}

\begin{lem}
Under \( \bar{P} \) the \( Y_{t} \) are \( {\cal N}\left( 0,1\right)  \)
i.i.d. random variables.
\end{lem}

\paragraph*{Proof}

The proof is based on the conditionnal Bayes'Theorem, it is a simple
rewriting of the Proof of Elliot , hence we have 

\[
\bar{P}\left( Y_{t+1}\leq \tau \left| {\mathcal{G}}_{t}\right. \right) =\bar{E}\left[ 1_{\{Y_{t+1}\leq \tau \}}\left| {\mathcal{G}}_{t}\right. \right] \]
 Thanks the conditionnal Bayes' Theorem we have : \[
\bar{E}\left[ 1_{\{Y_{t+1}\leq \tau \}}\left| {\mathcal{G}}_{t}\right. \right] \]
 \[
=\frac{E\left[ \Lambda _{t+1}1_{\{Y_{t+1}\leq \tau \}}\left| {\mathcal{G}}_{t}\right. \right] }{E\left[ \Lambda _{t+1}\left| {\mathcal{G}}_{t}\right. \right] }\]
 \[
=\frac{\Lambda _{t}}{\Lambda _{t}}\times \frac{E\left[ \lambda _{t+1}1_{\{Y_{t+1}\leq \tau \}}\left| {\mathcal{G}}_{t}\right. \right] }{E\left[ \lambda _{t+1}\left| {\mathcal{G}}_{t}\right. \right] }.\]
 Now \[
E\left[ \lambda _{t+1}\left| {\mathcal{G}}_{t}\right. \right] =\int _{-\infty }^{\infty }\frac{\left< \sigma ,X_{t}\right> \phi (Y_{t+1})}{\phi (\varepsilon _{t+1})}\times \phi (\varepsilon _{t+1})d\varepsilon _{t+1}\]
 \[
=\int _{-\infty }^{\infty }\left< \sigma ,X_{t}\right> \phi (F_{X_{t}}(Y^{t}_{t-p+1})+\left< \sigma ,X_{t}\right> \times \varepsilon _{t+1})d\varepsilon _{t+1}=1\]
 and since \( \varepsilon _{t+1}=\frac{Y_{t+1}-F_{X_{t}}\left( Y_{t-p+1}^{t}\right) }{\left\langle \sigma ,X_{t}\right\rangle } \)
: \[
\begin{array}{lcl}
\displaystyle \bar{P}\left( Y_{t+1}\leq \tau \left| {\mathcal{G}}_{t}\right. \right)  & = & E\left[ \lambda _{t+1}1_{\{Y_{t+1}\leq \tau \}}\left| {\mathcal{G}}_{t}\right. \right] \\
 &  & \\
 & = & \int _{-\infty }^{\infty }\frac{\left< \sigma ,X_{t}\right> \phi (Y_{t+1})}{\phi (\varepsilon _{t+1})}\times 1_{\{Y_{t+1}\leq \tau \}}\times \phi (\varepsilon _{t+1})d\varepsilon _{t+1}\\
 &  & \\
 & = & \int _{-\infty }^{\tau }\phi (Y_{t+1})dy_{t+1}=\bar{P}\left( Y_{t+1}\leq \tau \right) 
\end{array}\]
\( \blacksquare  \)

\subsection*{Proof of Theorem 2}

\begin{thm}
\label{theoreme_principal_elliott}Suppose \( H_{t} \) is a scalar
\( {\cal G} \)-adapted process of the form : \( H_{0} \) is \( {\cal F_{0}} \)
measurable, \( H_{t+1}=H_{t}+\alpha _{t+1}+\left\langle \beta _{t+1},V_{t+1}\right\rangle +\delta _{t+1}f\left( Y_{t+1}\right)  \),
\( k\geq 0 \), where \( V_{t+1}=X_{t+1}-AX_{t} \), \( f \) is a
scalar valued function and \( \alpha  \), \( \beta  \), \( \delta  \)
are \( {\cal G} \) predictable process (\( \beta  \) will be \( N \)-dimensional
vector process). Then : \begin{equation}
\label{equation_th_princip}
\begin{array}{lcl}
\displaystyle \gamma _{t+1}\left( H_{t+1}X_{t+1}\right)  & := & \gamma _{t+1,t+1}\left( H_{t+1}\right) \\
 & = & \sum _{i=1}^{N}\left\{ \left\langle \gamma _{t}\left( H_{t}X_{t}\right) ,\Gamma ^{i}\left( y_{t+1}\right) \right\rangle a_{i}\right. \\
 &  & +\gamma _{t}\left( \alpha _{t+1}\left\langle X_{t},\Gamma ^{i}\left( y_{t+1}\right) \right\rangle \right) a_{i}\\
 &  & +\gamma _{t}\left( \delta _{t+1}\left\langle X_{t},\Gamma ^{i}\left( y_{t+1}\right) \right\rangle \right) f\left( y_{t+1}\right) a_{i}\\
 &  & +\left( diag\left( a_{i}\right) -a_{i}a_{i}^{T}\right) \gamma _{t}\left( \beta _{t+1}\left\langle X_{t},\Gamma ^{i}\left( y_{t+1}\right) \right\rangle \right) 
\end{array}
\end{equation}
where \( a_{i}:=Ae_{i} \), \( a_{i}^{T} \) is the transpose of \( a_{i} \)
and \( diag\left( a_{i}\right)  \) is the matrix with vector \( a_{i} \)
for diagonal and zeros elsewhere. 
\end{thm}

\paragraph*{Proof}

Here again it is only a rewriting of the proof of Elliot. 

We begin with the two folowwing results :

\subparagraph{Result 1}

\begin{equation}
\label{lemme1}
\begin{array}{ll}
\bar{E}\left[ V_{t+1}\left| {\mathcal{Y}}_{t+1}\right. \right]  & =\bar{E}\left[ \bar{E}\left[ V_{t+1}\left| {\mathcal{G}}_{t},{\mathcal{Y}}_{t+1}\right. \right] \left| {\mathcal{Y}}_{t+1}\right. \right] \\
 & =\bar{E}\left[ \bar{E}\left[ V_{t+1}\left| {\mathcal{G}}_{t}\right. \right] \left| {\mathcal{Y}}_{t+1}\right. \right] =0.
\end{array}
\end{equation}

\subparagraph{Result 2}

\[
X_{t+1}X^{T}_{t+1}=AX_{t}(AX_{t})^{T}+AX_{t}V^{T}_{t+1}+V_{t+1}(AX_{t})^{T}+V_{t+1}V^{T}_{t+1}.\]
 Since \( X_{t} \) is of the form \( \left( 0,\cdots ,0,1,0,\cdots ,0\right)  \)
we have \[
X_{t+1}X^{T}_{t+1}=diag(X_{t+1})=diag(AX_{t})+diag(V_{t+1})\]
 so \[
V_{t+1}V^{T}_{t+1}=diag(AX_{t})+diag(V_{t+1})-A\, \, diag(X_{t})\, \, A^{T}-AX_{t}V^{T}_{t+1}-V_{t+1}(AX_{t})^{T}.\]
Finaly we obtain the result 

\begin{equation}
\label{lemme2}
\begin{array}{ll}
\left< V_{t+1}\right>  & :=E[V_{t+1}V^{T}_{t+1}\left| {\mathcal{F}}_{t}\right. ]\\
 & =E[V_{t+1}V^{T}_{t+1}\left| X_{t}\right. ]\\
 & =diag(AX_{t})-A\, \, diag(X_{t})\, \, A^{T}.
\end{array}
\end{equation}

\subparagraph{Main proff}

We have 

\[
\begin{array}{l}
\gamma _{t+1,t+1}(H_{t+1})=\bar{E}\left[ \bar{\Lambda }_{t+1}H_{t+1}X_{t+1}\left| {\mathcal{Y}}_{t+1}\right. \right] \\
=\bar{E}\left[ \left( AX_{t}+V_{t+1}\right) \left( H_{t}+\alpha _{t+1}+<\beta _{t+1},V_{t+1}>+\delta _{t+1}f(y_{t+1})\right) \times \bar{\Lambda }_{t+1}\left| {\mathcal{Y}}_{t+1}\right. \right] \\

\end{array}\]
 Thanks equation (\ref{lemme1}), \[
\begin{array}{l}
\gamma _{t+1,t+1}(H_{t+1})=\bar{E}\left[ \left( \left( H_{t}+\alpha _{t+1}+\delta _{t+1}f(y_{t+1})\right) AX_{t}+<\beta _{t+1},V_{t+1}>\right) \times \bar{\Lambda }_{t+1}\left| {\mathcal{Y}}_{t+1}\right. \right] .\\

\end{array}\]
 so : \[
\begin{array}{l}
\displaystyle \gamma _{t+1,t+1}(H_{t+1})=\sum _{j=1}^{N}\left\{ \bar{E}\left[ \left( \left( H_{t}+\alpha _{t+1}+\delta _{t+1}f(y_{t+1})\right) <AX_{t},e_{j}>e_{j}\right) \bar{\Lambda }_{t+1}\left| {\mathcal{Y}}_{t+1}\right. \right] \right\} \\
+\bar{E}\left[ <\beta _{t+1},V_{t+1}>\times \bar{\Lambda }_{t+1}\left| {\mathcal{Y}}_{t+1}\right. \right] ,
\end{array}\]
 hence \[
\begin{array}{l}
\displaystyle \gamma _{t+1,t+1}(H_{t+1})=\sum _{j=1}^{N}\sum _{i=1}^{N}\left\{ \bar{E}\left[ \left( \left( H_{t}+\alpha _{t+1}+\delta _{t+1}f(y_{t+1})\right) <X_{t},e_{i}>\right) \bar{\Lambda }_{t+1}a_{ji}e_{j}\left| {\mathcal{Y}}_{t+1}\right. \right] \right\} \\
+\bar{E}\left[ <\beta _{t+1},V_{t+1}>\times \bar{\Lambda }_{t+1}\left| {\mathcal{Y}}_{t+1}\right. \right] .
\end{array}\]
we have noted \( a_{i}=Ae_{i} \), so \[
\begin{array}{l}
\displaystyle \gamma _{t+1,t+1}(H_{t+1})=\sum _{i=1}^{N}\left\{ \bar{E}\left[ \left( \left( H_{t}+\alpha _{t+1}+\delta _{t+1}f(y_{t+1})\right) <X_{t},e_{i}>\right) \bar{\Lambda }_{t+1}a_{i}\left| {\mathcal{Y}}_{t+1}\right. \right] \right\} \\
+\bar{E}\left[ <\beta _{t+1},V_{t+1}>\times \bar{\Lambda }_{t+1}\left| {\mathcal{Y}}_{t+1}\right. \right] .
\end{array}\]
 Since for an adapted process \( H_{t} \) to the sigma-algebra \( {\mathcal{G}}_{t} \)
\[
\displaystyle \bar{E}\left[ \bar{\Lambda }_{t+1}H_{t}\left| {\mathcal{Y}}_{t+1}\right. \right] =\sum _{i=1}^{N}\left< \gamma _{t}(H_{t}X_{t}),\Gamma ^{i}(y_{t+1})\right> \]
 So, for all \( e_{r}\in \mathbb E \) \[
\begin{array}{ll}
\displaystyle \bar{E}\left[ \bar{\Lambda }_{t+1}H_{t}<X_{t},e_{r}>\left| {\mathcal{Y}}_{t+1}\right. \right]  & =\sum _{i=1}^{N}\left< \gamma _{t}(H_{t}X_{t}<X_{t},e_{r}>),\Gamma ^{i}(y_{t+1})\right> \\
 & =\sum _{i=1}^{N}\left< \gamma _{t}(H_{t}X_{t}X_{t}^{T}e_{r}),\Gamma ^{i}(y_{t+1})\right> \\

\end{array}\]
 But we have also : \[
\displaystyle \gamma _{t}(H_{t}X_{t}X_{t}^{T})=\sum _{i=1}^{N}\left< \gamma _{t}(H_{t}X_{t}),e_{i}\right> e_{i}e_{i}^{T},\]
 So we have : \[
\displaystyle \bar{E}\left[ \bar{\Lambda }_{t+1}H_{t}<X_{t},er>\left| {\mathcal{Y}}_{t+1}\right. \right] =\sum _{i=1}^{N}\left< \gamma _{t}(H_{t}X_{t}X_{t}^{T}e_{r}),\Gamma ^{i}(y_{t+1})\right> =\left< \gamma _{t}(H_{t}X_{t}),\Gamma ^{r}(y_{t+1})\right> .\]
 Since \( \alpha  \), \( \beta  \), \( \delta  \) are \( {\mathcal{G}} \)
predictible and \( f(y_{t+1}) \) mesurable with respect to \( {\mathcal{Y}}_{t+1} \),
the result (\ref{lemme2}) yield us the conclusion \( \blacksquare  \)

\end{document}